\documentclass[%
aip,
 amsmath,amssymb,
 reprint,%
]{revtex4-1}

\usepackage{graphicx}
\usepackage{dcolumn}
\usepackage{bm}

\usepackage[utf8]{inputenc}
\usepackage[T1]{fontenc}
\usepackage{mathptmx}
\usepackage{etoolbox}

\makeatletter
\def\@email#1#2{%
 \endgroup
 \patchcmd{\titleblock@produce}
  {\frontmatter@RRAPformat}
  {\frontmatter@RRAPformat{\produce@RRAP{*#1\href{mailto:#2}{#2}}}\frontmatter@RRAPformat}
  {}{}
}%
\makeatother
\begin{document}

\preprint{AIP/123-QED}

\title{High Precision, Low Excitation Capactiance Measurement Methods
  from 10 mK- to Room-Temperature}

\author{Lili Zhao}
\affiliation{International Center for Quantum Materials, Peking University, Beijing, China, 100871}
\author{Wenlu Lin}
\affiliation{International Center for Quantum Materials, Peking University, Beijing, China, 100871}
\author{Xing Fan}
\affiliation{College of Engineering and Applied Sciences, Nanjing University, Nanjing, China, 210093}
\author{Yuanjun Song}
\affiliation{Beijing Academy of Quantum Information Sciences, Beijing, China, 100193}
\author{Hong Lu}
\affiliation{College of Engineering and Applied Sciences, Nanjing University, Nanjing, China, 210093}
\author{Yang Liu}
\email{liuyang02@pku.edu.cn}
\affiliation{International Center for Quantum Materials, Peking University, Beijing, China, 100871}
\date{\today}

\begin{abstract}
  Capacitance measurement is a useful technique in studying quantum
  devices, as it directly probes the local particle charging
  properties, i.e. the system compressibility. Here we report one
  approach which can measure capacitance from mK to room temperature
  with excellent accuracy. Our experiments show that such a
  high-precision technique is able to reveal delicate and essential
  properties of high-mobility two-dimensional electron systems. 
\end{abstract}

\maketitle

\section{\label{sec:level1}Intorduction}

Capacitance contains useful information of electronic devices, as it
directly probes their electrical charging properties
\cite{PhysRevLett.21.212, PhysRevB.9.4410, PhysRevB.32.2696,
  MOSSER19865, PhysRevLett.68.3088, PhysRevB.47.4056,
  PhysRevB.50.1760}. Recently, the capacitance measurement at cryogenic
temperature has attracted significant attention in quantum studies and
has revealed a series of quantum phenomena \cite{PhysRevB.34.2995,
  PhysRevLett.78.4613, Hunt_NatureCom_2017, PhysRevLett.123.046601,
  PhysRevLett.68.674, nature23893, PhysRevLett.121.167601,
  PhysRevLett.122.116601, Irie_APE_2019}. Although high precision
capacitance measurement is widely used in studying classical devices
such as field effect transistors and diodes, it is extremely difficult
to be performed on quantum devices. Firstly, the quantum phenomena
usually emerge in fragile systems so that the excitation should be
sufficiently low to preserve the quantum properties of the
device. Secondly, many quantum devices can only be studied in
cryostats which host the low temperature and high magnetic field
environment. The meter-long cables connecting samples and room
temperature instruments have $\sim$ 100 pF capacitance, which is
orders-of-magnitude larger than the devices themselves. Thirdly, the
total power dissipation at the cryogenic sample stage must be limited
to sub-$\mu$W in order to maintain the low-temperature environment. This limits the use of active devices.

In order to reduce the crosstalk and signal leaking between cables,
many reported works use the cryogenic preamplifier to isolate the
input and output signals \cite{PhysRevLett.68.3088,
  Hunt_NatureCom_2017, PhysRevLett.123.046601, Hazeghi_rsi_2011,
  Verbiest_rsi_2019}.  Unfortunately, the preamplifiers usually
dissipate more than $\sim$10 $\mu$W heat at the sample stage, which
is sufficiently high to cause a noticeable temperature
raise. Sometimes, indirect probs of capacitance such as penetration
field, etc, are used which can provide qualitative information about quantum
phase transitions \cite{PhysRevB.50.1760, PhysRevLett.68.674,
  nature23893, PhysRevLett.121.167601, PhysRevLett.122.116601}.

In this report, we introduce a new approach for high precision
capacitance measurement from 10 mK- to room-temperature. We install a passive bridge
at the sample stage, and use a voltage-controlled-variable-resistance
to in-situ tune its balance. We use radio frequency excitation to
increase the output signal, and develop a high sensitivity redio frequency (RF) lock-in
technique to analyze the bridge output.

\section{Method}

\begin{figure*}[!htbp]
\includegraphics[width=0.95\textwidth]{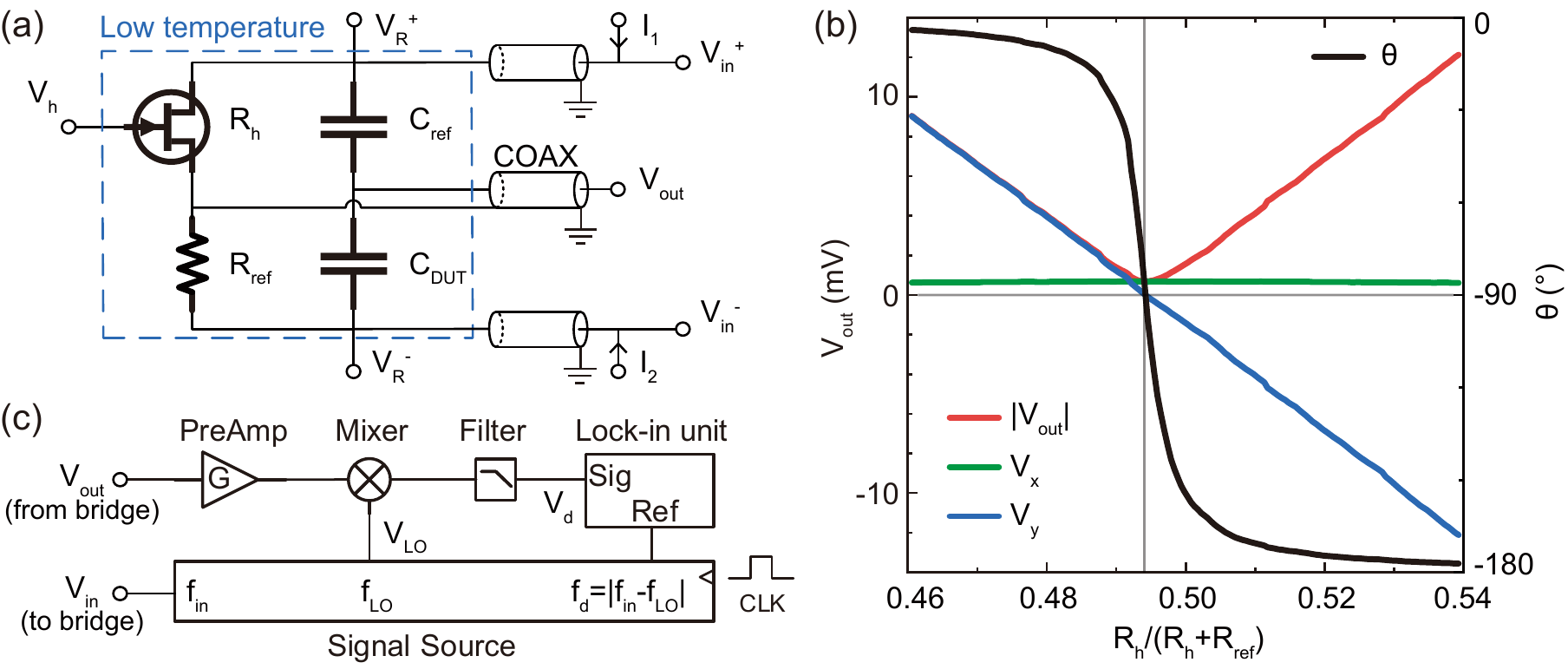}
\caption{\label{fig:fig1}Schematic diagram of our measurement setup. (a) The passive
  capacitance bridge installed at the sample stage consists
  $C_{\text{DUT}}$, $C_{\text{ref}}$, $R_{\text{ref}}$ and
  $R_{\text{h}}$. $R_{\text{h}}$ can be tuned by
  $V_{\text{h}}$. $V_{\text{in}}^+$ and $V_{\text{in}}^-$ are the two
  differentially coupled input signal, $V_{\text{out}}$ is the output
  signal. Coaxial cables are used to improve the transmission
  efficiency and minimize the parasitic capacitance. $I_1$ and $I_2$
  are two quasi-DC current inputs with different frequencies,
  $R_{\text{h}}$ ($R_{\text{ref}}$) can be deduced by analyzing the
  frequency component same as $I_1$ ($I_2$) of the differential
  voltages between $V_{\text{R}}^+$ and $V_{\text{R}}^-$. (b) One
  typical result of the "V-curve" procedure. The cross point is the
  balance point of the bridge, signaled by the $|V_{\text{out}}|$
  minimum, and $C_{\text{DUT}}$ can be calculated using
  Eq. (1). $V_{\text{out}}$ can be decomposed into $V_{\text{y}}$ and
  $V_{\text{x}}$. $V_{\text{x}}=0$ at the balancing point and depends
  on $\frac{R_{{\text{h}}}}{R_{\text{ref}}+R_{{\text{h}}}}$
  linearly. (c) Diagram of the radio-frequency lock-in setup. The
  bridge output $V_{\text{out}}$ passes through a room-temperature
  amplifier (PreAmp), a mixer (Mixer) and a low pass filter (Filter)
  and turns into a low frequency signal $V_{\text{d}}$ whose frequency
  is $f_{\text{d}}=|f_{\text{in}}-f_{\text{LO}}|$. $V_{\text{d}}$ is
  then analyzed by an audio-frequency digital lock-in unit.}
\end{figure*}

The passive bridge at the sample stage (Fig. 1(a)) consists of the
device under test $C_{\text{DUT}}$, the reference capacitor
$C_{\text{ref}}$, the reference resistor $R_{\text{ref}}$ and a
voltage-controlled-variable-resistor $R_{\text{h}}$. The excitation
voltage $V_{\text{in}}$ is differentially coupled to the
$V_{\text{in}}^+$ \& $V_{\text{in}}^-$ ports of the bridge by a RF transformer, and the
output signal $V_{\text{out}}$ is the voltage difference between the
midpoints of the capacitor and resistor arms. The bridge dissipates
only $\sim$ 10 nW when $V_{\text{in}}\simeq 1$ mV$_{\text{rms}}$,
mostly caused by $R_{\text{ref}}$ and $R_{\text{h}}$. We can tune the balance
of the bridge with $R_{\text{h}}$, which is the drain-to-source
resistance of a high electron mobility transistor (model
ATF35143). The bridge reaches it's balance point when the balance
condition \begin{equation}
  \label{eq:1}
\frac{C_{\text{DUT}}}{C_{\text{ref}}}=\frac{R_{\text{h}}}{R_{\text{ref}}}. 
\end{equation}
is achieved, signaled by the $|V_{\text{out}}|$ minimum (see
Fig. 1(b)). Two low-frequency current inputs $I_1$ and $I_2$ are
injected into the bridge. They flow into the ground through
$R_{\text{h}}$ and $R_{\text{ref}}$, respectively, and generate a
voltage difference
$(V_{\text{R}}^+-V_{\text{R}}^-)=I_1\cdot R_{\text{h}}-I_2\cdot
R_{\text{ref}}$. We can in-situ measure $R_{\text{h}}$ and
$R_{\text{ref}}$ simultaneously with lock-in technique by locking the
frequency of $(V_{\text{R}}^+-V_{\text{R}}^-)$ to be the same as $I_1$ and
$I_2$.  Therefore, as long as $C_{\text{ref}}$ is stable, the absolute
value of $C_{\text{DUT}}$ can be deduced by Eq.\ref{eq:1}. The
$|V_{\text{out}}|$ vs. $R_{\text{h}}$ curve is "V"-shaped, so we call
this procedure as the "V-curve" procedure for brevity.

The output impedance of this bridge is about
$\frac{1}{2\pi f (C_{\text{ref}}+C_{\text{DUT}})}$, where $f$ is the
frequency of $V_{\text{in}}$. $C_{\text{DUT}}$ and $C_{\text{ref}}$
are typically in the order of 0.1 pF, corresponding to output
impedance as large as $\sim$ 1 G$\Omega$ when $f = 1$ kHz. Therefore,
we increase $f$ to $\sim$ 100 MHz so that the output impedance
decreases to about 10 k$\Omega$. we use standard 50 $\Omega$ coaxial cable
and impedance matched RF preamplifier at room temperature for
broadband measurement. When the excitation amplitude is
$\simeq 1\text{mV}_{\text{rms}}$, $V_{\text{out}}$ is only about $\sim$
1 $\mu \text{V}_{\text{rms}}$ and 0.1\% change in $C_{{\text{DUT}}}$
corresponds to $\simeq 1$ nV$_{\text{rms}}$ variation in the
output. In order to analyze such a small signal, we combine the
superheterodyne and lock-in techniques (see Fig. 1(c)). The signal
source generates three single-frequency signals $V_{\text{in}}$,
$V_{\text{LO}}$ and $V_{\text{ref}}$ with frequencies $f_{\text{in}}$,
$f_{\text{LO}}$ and $f_{\text{d}}$, respectively, where
$f_{\text{d}}=|f_{\text{in}}-f_{\text{LO}}|$.  $V_{\text{in}}$ is sent
to the bridge as the voltage excitation. The bridge output
$V_{\text{out}}$ is amplified by a low-noise-amplifier, mixed with
$V_{\text{LO}}$ and low-pass-filtered, resulting in an audio frequency
signal $V_{\text{d}}$. We use a digital audio-frequency lock-in unit
to measure the amplitude $|V_{\text{d}}|$ and phase
$\theta_{\text{d}}$ in reference to $V_{\text{ref}}$, where
$|V_{\text{d}}|\propto |V_{\text{out}}|$ and $\theta_{\text{d}}$ is
the same as the phase difference between $V_{\text{out}}$ and
$V_{\text{in}}$ but differs by a constant. For simplicity, we quote
$|V_{\text{d}}|$ and $\theta_{\text{d}}$ as the amplitude
$|V_{\text{out}}|$ and the (relative) phase $\theta$ (from
$V_{\text{in}}$) of the bridge output.

\section{Calibration}

\begin{figure*}[!htbp]
\includegraphics[width=0.95\textwidth]{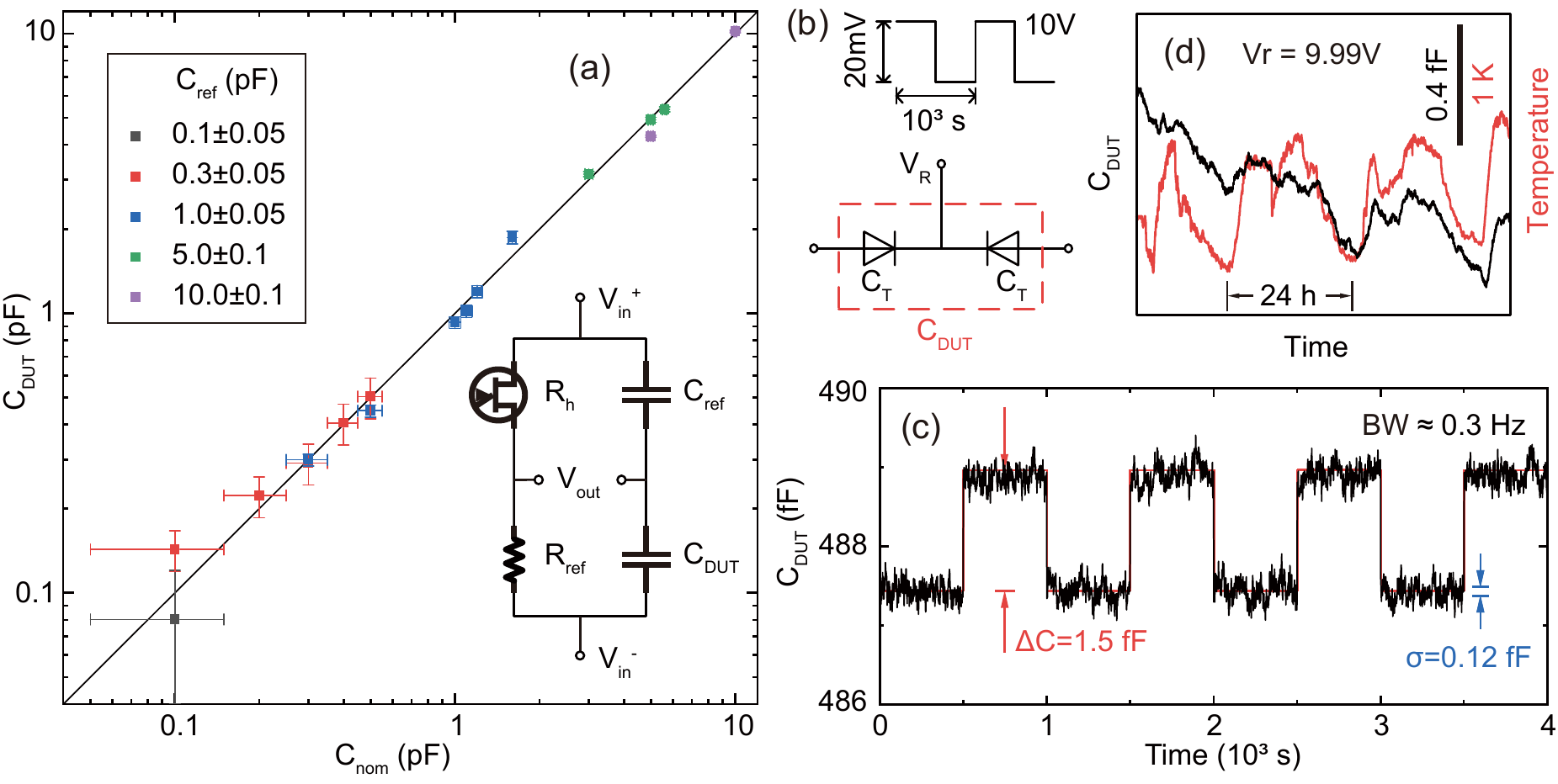}
\caption{Room temperature examination of our setup. (a)
  $C_{\text{DUT}}$ is the measured capacitors values using the
  "V-curve" procedure. $C_{\text{nom}}$ is their nominal value given
  by manufacturers, and the error bar is deduced from their
  tolerance. The black line is the ideal
  $C_{\text{DUT}}=C_{\text{nom}}$. Different colors of point data represent different $C_{\text{ref}}$ used. (b) Measuring two tunable
  capacitance diodes in series. The reverse voltage $V_{\text{R}}$
  tunes the value of each diode capacitance
  $C_{\text{T}}$. $V_{\text{R}}$ is a 1000-second-period, square wave
  with high/low voltages equal 10/9.98 V. (c) Measured
  $C_{\text{DUT}}$.  $C_{\text{DUT}}\approx487.5$ fF for
  $V_{\text{R}}=10 V$, and increases by $1.5$ fF when $V_{\text{R}}$
  decreases by 20 mV. The standard deviation is as low as 0.1 fF at
  0.3 Hz measurement bandwidth. (d) Long period monitoring of
  $C_{\text{DUT}}$ with $V_{\text{R}}$ fixed at 9.99 V (black), as
  well as the room temperature (red).}
\end{figure*}

We calibrate our setup at room temperature by measuring fixed-value capacitors whose
nominal value $C_{\text{nom}}$ ranges from 0.1 to 10 pF (see Fig.  2(a)). We select different $C_{\text{ref}}$
that is comparable with $C_{\text{DUT}}$. The y-axis is the measured
$C_{\text{DUT}}$ using the "V-curve" procedure with $\simeq$ 1.5
mV$_{\text{rms}}$ excitation amplitude and $\simeq 110$ MHz
frequency. The horizontal and vertical error bars are deduced from the
tolerance of $C_{\text{DUT}}$ and $C_{\text{ref}}$, which dominate the
inaccuracy at $<1$ pF. The measured $C_{\text{DUT}}$ matchs its
nominal value $C_{\text{nom}}$ even for capacitors as small as 0.1 pF,
evidencing that our setup is capable of measuring the absolute
capacitance value \footnote{In Fig.  2(a), $C_{{\text{DUT}}}$ slightly
  deviates from $C_{\text{nom}}$ when
  $C_{\text{DUT}} \ne C_{\text{ref}}$. This is not surprising because
  the bridge output is less sensitive to $C_{\text{DUT}}$ in theis case, so that the
  leaking signal causes more deviation.}.

The "V-curve" procedure can measure the absolute value of
$C_{\text{DUT}}$ with decent accuracy, however, continuously
monitoring $C_{\text{DUT}}$ while sweeping particular physical
parameter, such as magnetic field, temperature, etc, is of more
interest. We can deduce $C_{\text{DUT}}$ from $V_{\text{out}}$ using
the $V_{\text{out}}$ vs. $R_{\text{h}}$ relation obtained by the
"V-curve" procedure. At the vicinity of the balance point, the bridge
output $V_{\text{out}}$ is approximated as
\begin{equation}
  \label{eq:2}
  V_{\text{out}}\propto (\frac{C_{\text{DUT}}}{C_{\text{ref}}+C_{\text{DUT}}} - \frac{R_{{\text{h}}}}{R_{\text{ref}}+R_{{\text{h}}}})\cdot V_{\text{in}} + V_0,
\end{equation}
where $V_0$ represents the leaking signal. $V_{\text{out}}$ is a
vector and can be decomposed into two orthogonal scalar components
using its amplitude $|V_{\text{out}}|$ and phase $\theta$
\begin{equation}
  \label{eq:3}
  \left\{
    \begin{array}{lr}
      V_{\text{x}}=|V_{\text{out}}|\cdot \cos(\theta),\\
      V_{\text{y}}=|V_{\text{out}}|\cdot \sin(\theta),
    \end{array}\right.
\end{equation}
Note that, in the ideal case when $V_0=0$, $V_{\text{out}}$ changes
its sign, or $\theta$ changes by 180$^\circ$ as we tune the bridge
through its balance point. We set $\theta=-90^\circ$ at the balance
point so that $V_{\text{x}}=V_{\text{out}}$ is the expected bridge
output (Fig. 1(b)). In a careful measurement, leaking signal $V_0$
mostly comes from the capacitive coupling between the input and output
cables and has a $90^\circ$ phase shift from $V_{\text{in}}$. At the
vicinity of the balance point, $V_{\text{y}} \approx V_0$ and
$V_{\text{x}}$ has a linear dependence on
$\frac{R_{{\text{h}}}}{R_{\text{ref}}+R_{{\text{h}}}}$, see
Fig. 1(b). From the symmetry between the resistors and capacitors in
the bridge, we assume a single parameter, the sensitivity $S$, can
describe the dependence of $V_{\text{x}}$ on both $R_{\text{h}}$ and
$C_{\text{DUT}}$ by
\begin{equation}
  \label{eq:4}
  S=\frac{\partial V_{\text{x}}}{\partial \frac {R_{{\text{h}}}}{R_{\text{ref}}+R_{{\text{h}}}}} = \frac{\partial V_{\text{x}}}{\partial \frac {C_{\text{DUT}}}{C_{\text{ref}}+C_{\text{DUT}}}}.
\end{equation}
$S$ can be obtained by linearly fitting $V_{\text{x}}$ with
$\frac{R_{{\text{h}}}}{R_{\text{ref}}+R_{{\text{h}}}}$ while keep
$C_{\text{DUT}}$ fixed. Note that $S$ is inversely proportional to the
output impedance $1/2\pi f (C_{\text{ref}}+C_{\text{DUT}})$ and
corresponding corrections might be necessary. Thereafter, we can
deduce $C_{\text{DUT}}$ from $V_{\text{out}}$ by
\begin{equation}
  \label{eq:5}
  \frac {C_{\text{DUT}}}{C_{\text{ref}}+C_{\text{DUT}}} = \frac {R_{{\text{h}}}}{R_{\text{ref}}+R_{{\text{h}}}} - \frac{V_{\text{x}}}{S}. 
\end{equation}

Figs. 2(b-d) examine the feasibleness of the "monitoring"-mode. The
device-under-test is two back-to-back connected tunable capacitance
diodes (Infineon BB837) whose capacitance $C_{\text{T}}$ can be
controlled by the diode reverse voltage $V_{\text{R}}$. $V_{\text{R}}$
has a 9.99 V DC component which biases
$C_{\text{DUT}}=0.5 C_{\text{T}}$ to about 500 fF, as well as a 20
mV$_{\text{pp}}$, 1000-second-period square-wave AC component which induces a
small capacitance variation, see Fig. 2(b). The frequency and
amplitude of $V_{\text{in}}$ are 110 MHz and $\sim$1
mV$_{\text{rms}}$, respectively. We show the measured capacitance
using the "monitoring"-mode in Fig. 2(c). The measured
$C_{\text{DUT}}$ is about 487.5 fF when $V_{\text{R}}$=10 V, and
increases by 1.5 fF when $V_{\text{R}}$ decreases by 20mV; consistent
with the BB837 datasheet. The standard deviation of $C_{\text{DUT}}$
is about 0.12 fF at 0.3 Hz measurement bandwidth \footnote{The
  equivalent voltage noise of the whole setup at the input of the RF
  preamplifier has a spectrum density of about 8
  $\text{nV}/\sqrt{\text{Hz}}$.}. In short, this method can resolve
$\lesssim$ 240 ppm variation of a $\simeq$ 0.5 pF capacitor within a
few second. Fig. 2(d) shows $C_{\text{DUT}}$ measured for 72 hours,
which drifts by $<$ 1 fF in 72 h. The variation of $C_{\text{DUT}}$ is
in good agreement with the room temperature fluctuation, likely
introduced by the device itself.

\section{Measurements at mK-temperature}

\begin{figure}[!htbp]
\includegraphics[width=0.45\textwidth]{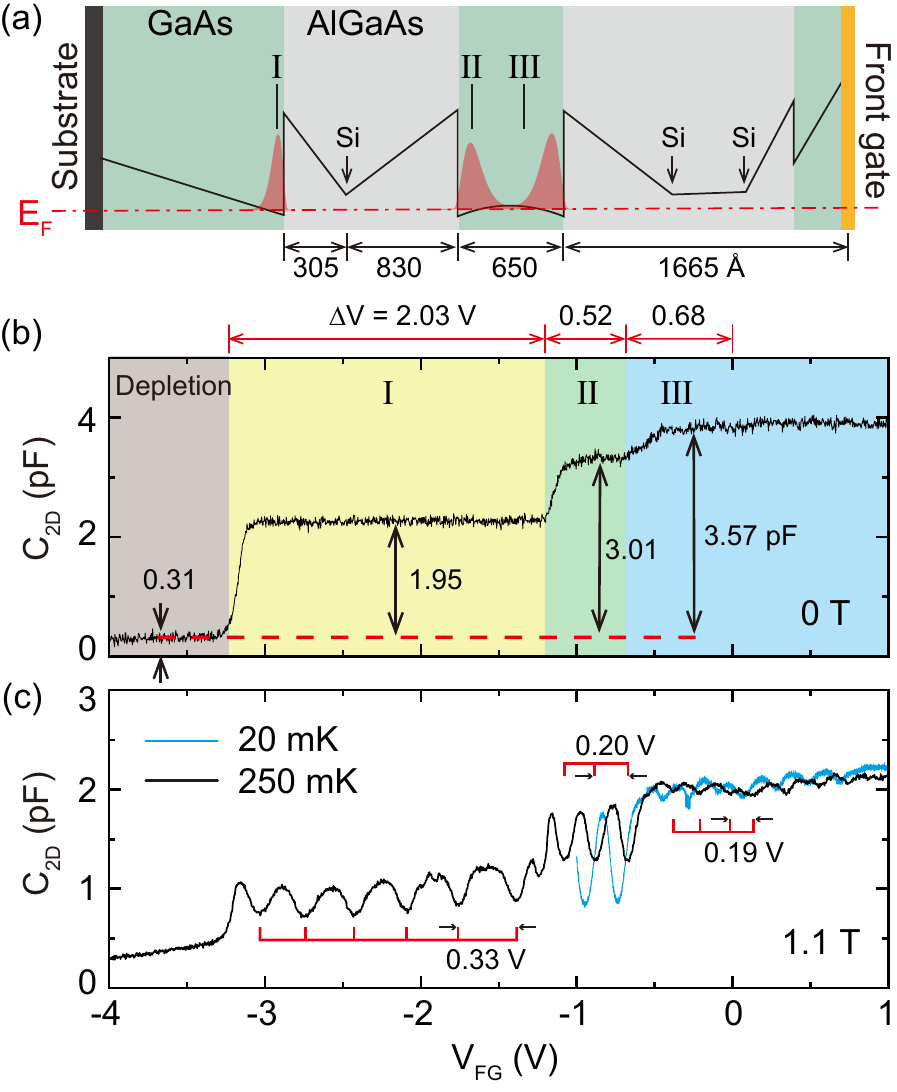}
\caption{ Studying high-mobility 2DEG in GaAs/AlGaAs heterostructure
  samples with capacitance measurement.  (a) Sample structure. Black,
  green, gray and yellow regions represents the substrate, the GaAs
  layer, the AlGaAs layer and the metallic front gate,
  respectively. The black solid line represents the conduction band
  edge and the red dash-dot line is the Fermi level
  $E_{\text{F}}$. The vertical arrow indicates the $\delta$-doping
  positions. The red shaded areas represent the charge distribution of
  2DEG at zero gate voltage. I, II and III indicate the effective
  charge position of the top-most 2DEG in the three cases of (b).  (b)
  The measured gate-to-2D capacitance $C_{\text{2D}}$. The four
  $C_{\text{2D}}$ plateaus indicate four different working senario of
  the device: Depletion, Cases I, II and III. The black and red
  double-arrow lines marks the capacitance value and the voltage range
  for each plateau. (c) $C_{\text{2D}}$ vs. $V_{\text{FG}}$ at 1.1 T
  perpendicular magnetic field. The red lines mark the periods of
  oscillation of the 250-mK trace. }
\end{figure}

We install the capacitance setup into an Oxford Triton 400 dilution
refrigerator, and study the gate-to-2D capacitance of
a high-mobility two-dimensional electron gas (2DEG) sample at mK
temperature in Fig. 3 \footnote{The mixing chamber temperature
  increases from $\simeq 8$ mK to $\simeq 17$ mK after we install
  three semi-rigid coaxial cables. In a separate report, we use
  extra-thin flexible coaxial wires and the mixing chamber temperature can
  be kept below 10 mK during the measurement
  \cite{Zhao_arxiv_2022}.}. The sample consists an AlGaAs/GaAs/AlGaAs
quantum well structure grown by molecular beam epitaxy, as illustrated
in Fig. 3(a). The mobility is about 2 $\text{m}^2/(\text{V} \cdot \text{s})$. The 650-$\text{\AA}$-wide GaAs quantum well resides 1665
$\text{\AA}$ below the surface, bound by AlGaAs spacer-layers on both
sides. We grow three Si-$\delta$-doping-layers in the spacer-layers,
two in the surface-side one, and one in the substrate-side one. At
zero front gate bias, a bilayer-like 2DEG forms inside the quantum
well, and another 2DEG forms at the interface between the
substrate-side AlGaAs spacer-layer and the GaAs buffer-layer, see Fig.
3(a). We evaporate two concentric gates and measure the gate-to-gate
capacitance, i.e. the two gate-to-2D capacitors which are
serial-connected by the 2DEG. The inner gate radius is 60
$\mu\text{m}$, and the gap between the two gates is 20 $\mu\text{m}$.
The outer-gate is much bigger than the inner-gate, so that
$C_{\text{DUT}}$ is approximately the inner-gate-to-2D capacitance. An isolation capacitor \footnote{The isolation capacitor is 22 nF in this measurement, which needs to be much larger than the device capacitance} is used between the inner-gate
and the bridge so that a DC gate bias $V_{\text{FG}}$ can be
applied. The frequency and amplitude of $V_{\text{in}}$ is 17 MHz and
$\sim$0.4 mV$_{\text{rms}}$, respectively. We don't see any increase
of the mixing chamber temperature ($\lesssim$ 17 mK) when turn on the
measurement, evidencing that the heat load at the sample stage is
negligible \cite{Note3}.

\begin{table*} \caption{\label{tab:table1}The parameters of 2DEG
    learnt by our measurement. The subscript $i=1,2,3$ indicates the
    parameter in cases I, II and III, respectively.}
\begin{ruledtabular}
{\begin{tabular}{cccccc}
Cases & $C_{\text{DUT}}^i$ (pF) & $d_i$ ($\text{\AA}$) & $\Delta V_i$ (V) & $n_i$ ($10^{11} cm^{-2}$) & $V_B^i$ (V) \\
\hline
I        & 1.95        & 3500            & 2.03 & 4.17        & 0.33\\
II       & 3.01        & 2273            & 0.52 & 1.64        & 0.20\\
III      & 3.57        & 1913            & 0.68 & 2.55        & 0.19\\
I:II:III & 1:1.54:1.83 & 1/(1:1.54:1.83) & $\backslash$ & 1:0.39:0.61 & 1/(1:1.65:1.74)    \\
\end{tabular}}
\end{ruledtabular}
\end{table*}

Fig. 3(b) shows the $C_{\text{DUT}}$ as a function of $V_{\text{FG}}$
measured at 250 mK \footnote{In this manuscript, all quoted
  temperatures are mixture chember plate temperatures of the dilution
  refrigerator}.  The electron charging falls into four scenarios,
corresponding to the four plateaus seen in the Fig. 3(b)
data. Depletion: The 2DEG under the inner-gate is completely depleted
when $V_{\text{FG}}$ is small and the $C_{\text{DUT}}$ is close to
zero. Case I: As $V_{\text{FG}}$ increases, the first 2DEG starts to
form at the interface between the buffer-layer and substrate-side
spacer-layer. Case II: The second 2DEG forms at the substrate-side
interface of the quantum well when $V_{\text{FG}}$ reaches -1.20
V. Case III: Keep increasing $V_{\text{FG}}$ increases the charge
density inside the quantum well as well as makes it more
symmetric. The effective charging position becomes close to the
quantum well center when electrons starts to fill the second subband,
whose charge distribution locates at the surface-side interface of the
quantum well.

The $C_{\text{DUT}}= 0.31$ pF in the depletion region is likely the
stray capacitance of the inner gate. We subtract it from the measured
capacitances and summarize the plateau heights corresponding to cases I,
II, and III in Table I as $C_{\text{DUT}}^i$, $i=1,2,3$ (see also Fig. 3(b)). Note
that $C_{\text{DUT}}^i$ is inversely proportional to $d_i$, the distance between
the surface and the effective position of the topmost 2DEG
\cite{PhysRevB.47.4056}. $d_1\approx 3500 \text{\AA}$ because the
effective position of a heterojunction 2DEG is estimated to be
$\simeq$50 $\text{\AA}$ away from the interface
\cite{RevModPhys.54.437}. We then calculate $d_2$ and $d_3$ using the
relation $C_{\text{DUT}}^i/C_{\text{DUT}}^j=d_j/d_i$ and list them in Table I. These results
suggest that the effective position of the quantum-well 2DEG is about
42 $\text{\AA}$ ($=2315 \text{\AA}-d_2$) away from its
substrate-side interface when it first appears, and moves to 248
$\text{\AA}$ ($=d_1-1665 \text{\AA}$) away from its surface-side
interface when electron occupies the second-subband. A comprehensive
investigation may track this evolution more thoroughly, which is out
of our scope here.

When subjected into a large perpendicular magnetic field $B$, the
electrons of a 2DEG are quenched into discrete Landau levels, leading
to density-dependent compressibility. When we tune the 2DEG density by
sweeping $V_{\text{FG}}$ (see Fig. 3(c)), we observe a $C_{\text{DUT}}$
oscillation where each period corresponds to the occupation of two
Landau levels (two spins) in the topmost 2DEG \cite{Irie_APE_2019}. we
extracted the oscillation period $V_B^i$ for the three cases and list
them in Table I. The reciprocal ratio of $V_B^i$ is similar to the
ratio of $C_{\text{DUT}}^i$, consistent with our expectation that $C_{\text{DUT}}^i$ is
inversely proportional to $V_B^i$ as
$C_{\text{DUT}}^i\propto \frac {2 Be^2}{h}/V_B^i $, where $h$ is the Planck's
constant, $e$ is the electron charge and $B$ is the magnetic
field. Note that the capacitance oscillation is strong in Case I and
II, but heavily damped in Case III. At lower temperature 20 mK, the
oscillation becomes more pronounced in case II, but remain roughly
unchanged in case III. This is possibly because the two subbands have
spatially separated charge distribution which smooths the
compressibility oscillation.

We can also calculate the density of each 2DEG using the relation
$n_i=\frac {\varepsilon_{\text{r}}
  \varepsilon_0}{e}\cdot \Delta V_i/d_i$, where
$\varepsilon_{\text{r}}=13$ is the relative dielectric constant,
$\varepsilon_0$ is the permittivity of vacuum and $\Delta V_i$ is the
width of the plateau in the Case-$i$ region, see Fig. 3(b). We find that
$n_2/n_1=0.39$ agrees with the reciprocal ratio of the distance from
the doping layer to the corresponding 2DEG (355
$\text{\AA} : 872 \text{\AA} =0.41$, see Fig.  3(a)).

\section{Conclusion}

We have introduced a high-precision, low-excitation capacitance
measurement method for 10 mK- to room-temperature experiments. We are able to
measure the absolute capacitance value using the "V-curve" procedure,
and monitor the variation of capacitor using $V_{\text{out}}$. With
about 1 mV$_{\text{rms}}$ excitation voltage, we can resolve 240 ppm
variation of a 500 fF capacitor. We measure the gate-to-2D capacitance
of a high-mobility two-dimensional electron system at mK-temperature
and extract consistent, essential information of the device. The
results demonstrate that our capacitance bridge can detect extremely
small capacitance fluctuation with mV-excitation at 10 mK-temperature.

\begin{acknowledgments}
  Y. Liu acknowledge support by the National Basic Research Program of China (Grant No. 2019YFA0308403) and the National Natural Science Foundation of China (Grant No. 92065104 and 12074010)
  for sample fabrication and measurement. H. Lu acknowledges the support from the National Key R\&D Program of China (2018YFA0306200) and the National Natural Science Foundation of China (Grant No. 51732006) for material growth. We thank M. Shayegan, Lloyd
  Engle, Jianhao Chen and Xi Lin for valuable discussion.
\end{acknowledgments}

\bibliography{bib_full}

\end{document}